\documentclass[sigconf]{acmart}

\newcommand{\figref}[1]{Fig.~\ref{#1}}
\newcommand{\tabref}[1]{Table~\ref{#1}}
\renewcommand{\eqref}[1]{Eq.~(\ref{#1})}
\usepackage{textcomp} 
\usepackage{multirow} 
\usepackage{subcaption}
\usepackage{pifont} 
\newcommand{\cmark}{\ding{51}}
\newcommand{\xmark}{\ding{55}}
\usepackage{enumitem}

\AtBeginDocument{%
  \providecommand\BibTeX{{%
    \normalfont B\kern-0.5em{\scshape i\kern-0.25em b}\kern-0.8em\TeX}}}

\setcopyright{rightsretained} 
\copyrightyear{2023} 
\acmYear{2023} 
\acmConference{SIGGRAPH '23 Posters}{August 06-10, 2023}{Los Angeles, CA, USA}
\acmBooktitle{Special Interest Group on Computer Graphics and Interactive Techniques Conference Posters (SIGGRAPH '23 Posters), August 06-10, 2023}
\acmDOI{10.1145/3588028.3603650}
\acmISBN{979-8-4007-0152-8/23/08}

\begin{document}

\title[Point Anywhere]{Point Anywhere: Directed Object Estimation from Omnidirectional Images}


\author{Nanami Kotani}
\affiliation{
  \institution{Tokyo Institute of Technology}
  \city{Tokyo}
  \country{Japan}
}
\email{kotani.titech@gmail.com}

\author{Asako Kanezaki}
\affiliation{
  \institution{Tokyo Institute of Technology}
  \city{Tokyo}
  \country{Japan}
}
\email{kanezaki@c.titech.ac.jp}





\renewcommand{\shortauthors}{N. Kotani and A. Kanezaki}

\begin{abstract}
  One of the intuitive instruction methods in robot navigation is a pointing gesture. 
  In this study, we propose a method using an omnidirectional camera to eliminate the user/object position constraint and the left/right constraint of the pointing arm. 
  Although the accuracy of skeleton and object detection is low due to the high distortion of equirectangular images, the proposed method enables highly accurate estimation by repeatedly extracting regions of interest from the equirectangular image and projecting them onto perspective images. 
  Furthermore, we found that training the likelihood of the target object in machine learning further improves the estimation accuracy.
\end{abstract}


\begin{teaserfigure}
  \includegraphics[width=\textwidth]{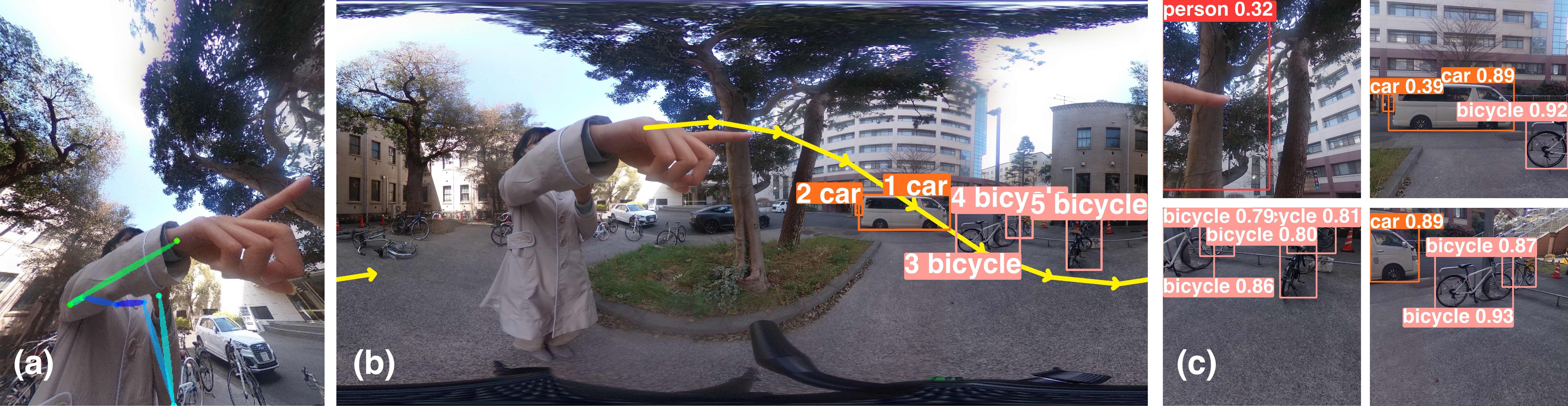}
  \caption{Overview of the proposed method. (a) Human skeleton is detected in a projected perspective image. (b) A great circle in the pointing direction is estimated by back-projecting the human skeleton. (c) Objects are detected in several regions of interest on the great circle projected to perspective images. Object regions are then back projected to (b) the equirectangular image to calculate the distances to the great circle. The ground truth object in this case is the first car.}
  \Description{The equirectangular image of a user pointing to a white car outdoors.}
  \label{fig:overview}
\end{teaserfigure}

\maketitle

\section{Introduction}
In recent years, voice-recognition devices such as smart speakers have been attracting attention. However, it is difficult to give instructions quickly because the instructions need to be verbalized.
Combining pointing object estimation from images with spoken instructions could be an effective solution to this problem. Pointing object estimation is also useful for indicating the destination of an autonomous mobile robot. 
Many methods have been proposed for estimating the pointing position using ordinary cameras \cite{Hu_2010_ICPR, Jaiswal_2018_UPCON, Azari_2019_CRV}, but the narrow field of view of cameras limits the standing position of users. 
Recently, an omnidirectional camera-based pointing position estimation method has been studied \cite{Shiratori_2021_ICIC}. In this method, all CG images used to train the pointing position estimation network point to the wall with the right hand. Additional training may be required for the network when the pointing position is other than a wall or for pointing with the left hand.

In this research, the goal is the pointing object estimation to provide simple and intuitive user instructions. By using omnidirectional images, users can give instructions without being aware of their standing position or the location of the target object in a wide range of areas with either the left or right arm.

\section{Approach}
\begin{figure*}[t!]
    \begin{center}
    \begin{minipage}{.48\linewidth}
        \includegraphics[width=\linewidth]{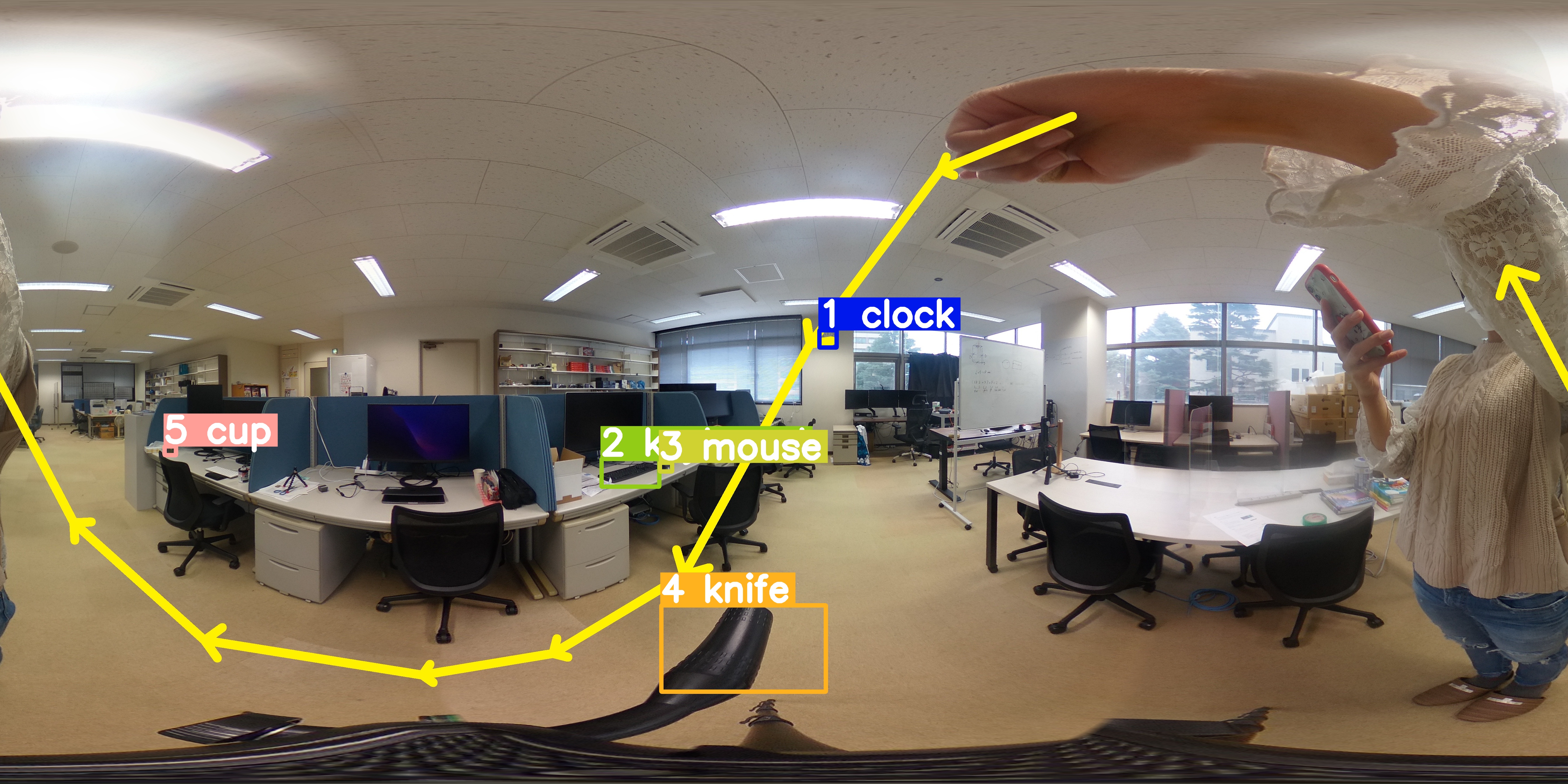}
        \subcaption{The ground truth is the first clock.}
        \label{fig:success1}
    \end{minipage}
    \begin{minipage}{.48\linewidth}
        \includegraphics[width=\linewidth]{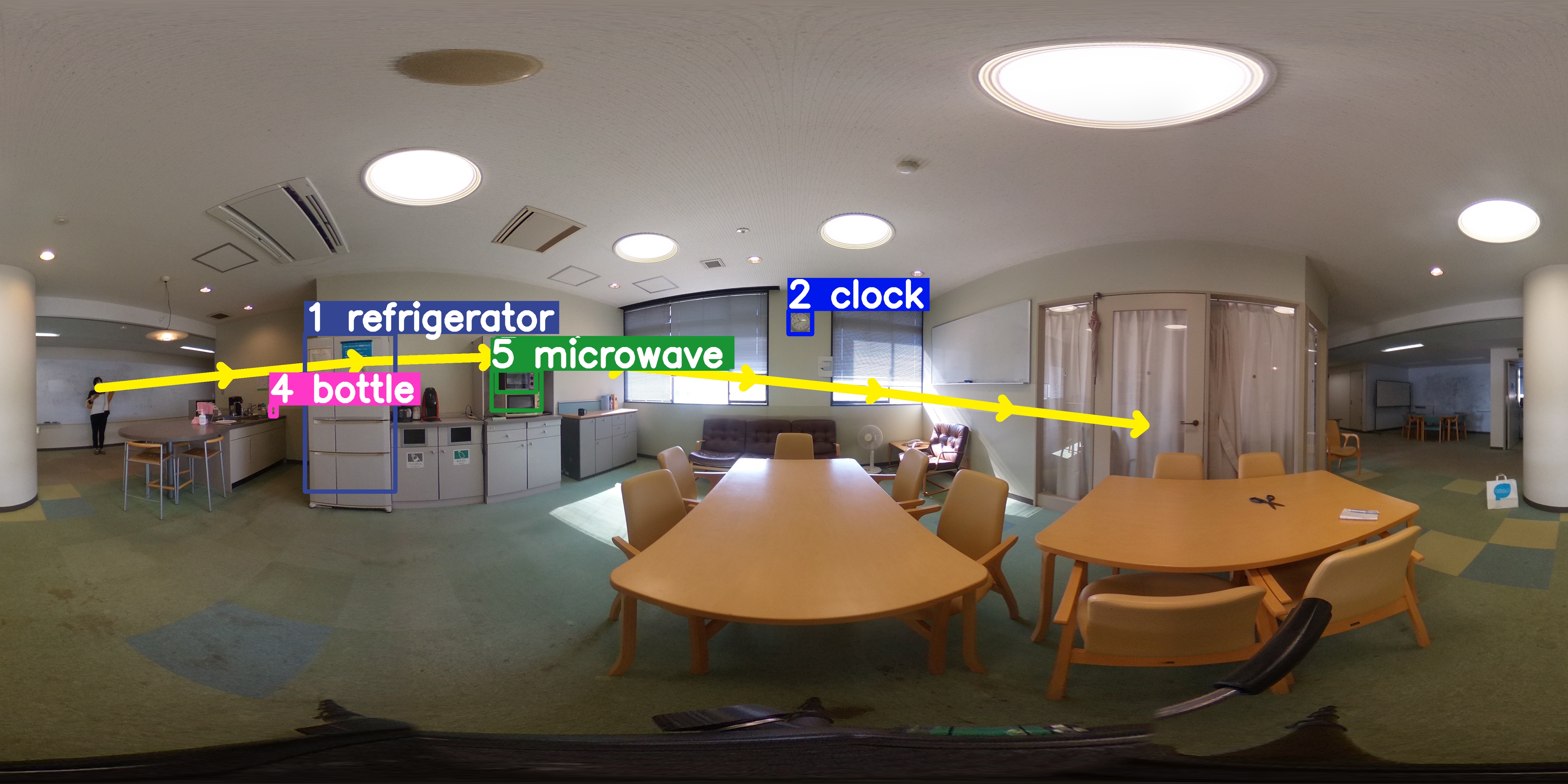}
        \subcaption{The ground truth is the first refrigerator.}
        \label{fig:success2}
    \end{minipage}
    \caption{Successful examples of the proposed method}
    \Description{(a) The user points to the clock on the wall. (b) The user points to the white refrigerator. }
    \end{center}
\end{figure*}

An overview of the proposed method is given in \figref{fig:overview}a-c. We have focused on the fact that a natural pointing gesture in daily life often involves pointing to a specific object, rather than to an empty space. Therefore, in this study, we deal with the case where the user is pointing to a certain object. 
As can be seen from \figref{fig:overview}b, the equirectangular image is distorted because the spherical surface is represented as a plane. Therefore, a pointing vector that is a straight line in real space becomes a great circle in the omnidirectional image. Since the estimation of the pointing direction is based on 2D vectors on the image, there is a problem of not being able to determine how far to extend the pointing vector. 
In this study, we address this problem using object detection and selection.
The proposed method consists of three processes.

The first process is the estimation of the pointing direction using the skeleton of the user. 
First, person detection is performed by YOLOv5\footnote{\url{https://github.com/ultralytics/yolov5}} on an equirectangular image. The longitude and latitude of the user are calculated from the bounding box. Next, skeletal detection is performed using OpenPose \cite{Cao_2017_CVPR} on a perspective image in which only the area around the user is projected.
Then the arm with the elbow extended \cite{Herbort_2018_PR} and located at the top of the image \cite{Azari_2019_CRV} is detected 
as the pointing arm. In this study, the pointing vector is the average of the shoulder and fingertip vectors and the head and fingertip vectors. The pointing vector is then fitted to a great circle to extend to the spherical space. The amplitude and phase of the great circle varies with the angle between the arm and the camera.

Next, regions of interest in the directional space are projected onto the perspective images. 
The regions of interest are extracted by sliding the great circle in the direction of the pointing vector at 30\textdegree ~ longitude intervals.
This prevents objects just between the region of interest from failing object detection, since the viewing angle of the perspective image is 60\textdegree.
Then a set of candidate objects is obtained by applying YOLOv5 to each perspective image.

Finally, the object regions on the obtained perspective images are projected onto the equirectangular image. 
The distance between the pointing vector and the object region is then calculated. 
There are multiple objects in the great circle in most cases, so we propose object selection methods.
The pointing object is selected mainly based on the distance. 
Let $o_i$ be the $i$th detected object instance and $S$ be the total number of detected pointing object candidates.
\begin{itemize}[leftmargin=*]
    \item $d_i$: Distance between the pointing vector and the center of $o_i$
    \item $q_i$: Number of objects belonging to the category $o_i$ belongs to
    \item $l_i = q_i / S$: Frequency of occurrence of the category $o_i$ belongs to
    \item $c_i$: Confidence of object detection for $o_i$
    \item $a_i$: Area of object region of $o_i$
    \item $h_i$: Horizontal distance between $o_i$ and the user
\end{itemize}
A two-class classification of whether a candidate is the pointing object is performed using linear SVC. Five variables are used as explanatory variables: $d_i$, $l_i$, $c_i$, $a_i$, and $h_i$. The data is standardized so that the mean is 0 and the variance is 1.

\begin{table}[t!]
  \caption{TOP-1 Accuracy}
  \vspace{-3mm}
  \label{tab:top1_acc}
  \begin{tabular}{llccc}
    \toprule
    \multicolumn{2}{l}{Projection in skeletal detection}&\xmark&\cmark&\cmark \\
    \multicolumn{2}{l}{Projection in object detection}&\xmark&\xmark&\cmark \\
    \midrule
    \multirow{2}{*}{Object selection}&Distance to vector&0.07&0.11&0.19 \\
    &Linear SVC&0.15&0.16&\textbf{0.27} \\
    \bottomrule
\end{tabular}
\end{table}

The quantitative evaluation is shown in \tabref{tab:top1_acc}.
We used $110$ images for training the linear SVC and $180$ images for testing.
A total of $4,884$ objects in $22$ categories were detected in $290$ images.
Successful examples are shown in \figref{fig:success1} and \figref{fig:success2}. The yellow arrows in the figures represent the instruction vectors fitted to the great circles. The number to the left of the object category name indicates the object's top estimated rank.

\section{Conclusion}
We proposed a method for estimating a pointing object from an omnidirectional image. Since the distortion of omnidirectional images is large, projecting only the region of interest onto the perspective image improves the accuracy of the estimation of the pointing object.
We also showed that the object selection performance was improved by learning the metrics using linear SVC.



\bibliographystyle{ACM-Reference-Format}
\bibliography{ref}

\end{document}